\documentclass[preprint,prb,aps,floats,amssymb,showpacs,draft,%
floatfix,superscriptaddress]{revtex4} %,nofootinbib
\usepackage{psfig}

\begin{document}

\title{ 
Multicritical Nishimori point in the phase diagram of the 
$\pm J$ Ising model on a square lattice
} 

\author{Martin Hasenbusch} 
\affiliation{ 
Institut f\"ur Theoretische Physik, Universit\"at Leipzig, 
Postfach 100 920, D-04009 Leipzig, Germany.
} 
\author{Francesco Parisen Toldin} 
\affiliation{ 
Max-Planck-Institut f\"ur Metallforschung,
Heisenbergstrasse 3, D-70569 Stuttgart, Germany\\
and Institut f\"ur Theoretische und Angewandte Physik,
Universit\"at Stuttgart,
Pfaffenwaldring 57, D-70569 Stuttgart, Germany.
} 
\author{Andrea Pelissetto} 
\affiliation{Dipartimento di Fisica
  dell'Universit\`a di Roma ``La Sapienza" and INFN, \\
  Piazzale Aldo Moro 2, I-00185 Roma, Italy.}
\author{Ettore Vicari} 
\affiliation{ 
Dipartimento di Fisica dell'Universit\`a di Pisa and INFN,\\
Largo Pontecorvo 3, I-56127 Pisa, Italy.  } 

\date{\today}

\begin{abstract}
  We investigate the critical behavior of the random-bond $\pm J$ Ising model 
  on a square lattice at the multicritical Nishimori point
  in the $T$-$p$ phase diagram,
  where $T$ is the temperature and $p$ is the disorder parameter
  ($p=1$ corresponds to the pure Ising model).  We perform a finite-size
  scaling analysis of high-statistics Monte Carlo simulations along the
  Nishimori line defined by $2p-1={\rm Tanh}(1/T)$, along which the 
  multicritical point lies. The multicritical Nishimori point is located at
  $p^*=0.89081(7)$, $T^*=0.9528(4)$, and the renormalization-group dimensions
  of the operators that control the multicritical behavior are $y_1=0.655(15)$
  and $y_2 = 0.250(2)$; they correspond to the thermal exponent
  $\nu\equiv 1/y_2=4.00(3)$ and to the crossover exponent 
  $\phi\equiv y_1/y_2=2.62(6)$.
\end{abstract}

\pacs{75.10.Nr, 64.60.Fr, 75.40.Cx, 75.40.Mg}

%% 75.10.Nr Spin-glass and other random models
%% 75.40.Cx Static properties (order parameter, static susceptibility, 
%%          heat capacities, critical exponents, etc.)
%% 75.40.Mg Numerical simulation studies
%% 64.60.Fr Equilibrium properties near critical points, critical exponents

\maketitle

% ========================= BODY =========================
%\narrowtext

\section{Introduction}

The $\pm J$ Ising model on a square lattice represents an interesting
theoretical laboratory, in which one can study the effects of quenched
disorder and frustration on the critical behavior of 
two-dimensional (2D) spin systems.  It
is defined by the lattice Hamiltonian
\begin{equation}
{\cal H} = - \sum_{\langle xy \rangle} J_{xy} \sigma_x \sigma_y,
\label{lH}
\end{equation}
where $\sigma_x=\pm 1$, the sum is over pairs of nearest-neighbor sites of a
square lattice, and the exchange interactions $J_{xy}$ are uncorrelated
quenched random variables, taking values $\pm J$ with probability distribution
\begin{equation}
P(J_{xy}) = p \delta(J_{xy} - J) + (1-p) \delta(J_{xy} + J). 
\label{probdis}
\end{equation}
In the following we set $J=1$ without loss of generality.  For $p=1$ we
recover the standard Ising model, while for $p=1/2$ we obtain the bimodal
Ising spin-glass model.  The $\pm J$ Ising model is a simplified model
\cite{EA-75} for disordered spin systems showing glassy behavior in some
region of their phase diagram.  The random nature of the short-ranged
interactions is mimicked by nearest-neighbor random bonds.  The 2D $\pm J$
Ising model is also interesting for the description of quantum Hall
transitions,\cite{CF-97,GRL-01,CRKHAL-01} and for its applications in coding
theory.~\cite{WHP-03,Kitaev-03,DKLP-02,Nishimori-01}

\begin{figure*}[tb]
\centerline{\psfig{width=9truecm,angle=0,file=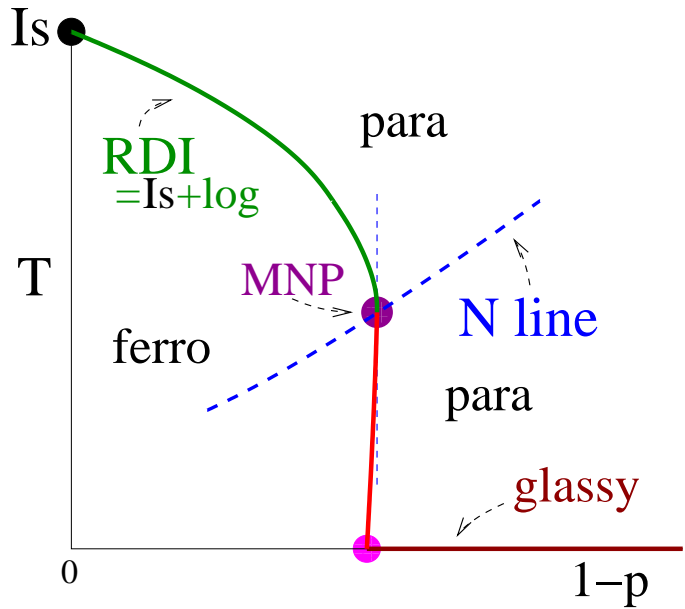}}
\caption{(Color online)
Phase diagram of the square-lattice $\pm J$ Ising model in the $T$-$p$
plane. }
\label{phdia}
\end{figure*}

The $T$-$p$ phase diagram of the 2D $\pm J$ Ising model is sketched in
Fig.~\ref{phdia} (it is symmetric for $p\rightarrow 1-p$ and thus we only
report it for $1-p<1/2$).  It has been investigated and discussed in several
works, see, e.g.,
Refs.~\onlinecite{CF-97,WHP-03,Nishimori-01,%%
  Nishimori-07,PHP-06,Queiroz-06,AH-04,KR-03,QS-03,MNN-03,MC-02,%%
  NN-02,HPP-01,Nobre-01,AQS-99,OI-98,BGP-98,MB-98,KR-97,SA-96,%%
  ON-93,Kitatani-92,ON-87,Nishimori-86}.  For sufficiently small values of 
$1-p$, which is the
probability of antiferromagnetic bonds, the model presents a paramagnetic
phase and a ferromagnetic phase, separated by a transition line.  The
paramagnetic-ferromagnetic (PF) transition line starts at the Ising point
$X_{\rm Is}=(T=T_{\rm Is},p=1)$, where $T_{\rm
  Is}=2/\ln(1+\sqrt{2})=2.26919...$ is the critical temperature of the 2D
Ising model, and extends up to the multicritical Nishimori point (MNP) at
$X_{\rm MNP}=(T^*,p^*)$, with $T^*\approx 0.95$ and $p^*\approx 0.89$.  Along
this line, the critical behavior is analogous to that observed in 
2D randomly dilute 
Ising (RDI) models.\cite{Shalaev-84,Shankar-87,BFMMPR-97,HPPV-08}
It is controlled by the pure Ising fixed point and 
disorder is marginally irrelevant, giving
rise to universal logarithmic corrections, as shown in
Refs.~\onlinecite{LC-87,HPPV-08}.  As argued in
Refs.~\onlinecite{GHDB-85,LH-88,LH-89}, the MNP is located along the so-called
Nishimori line ($N$ line)~\cite{Nishimori-81,Nishimori-01} defined by the
equation
\begin{equation}
{\rm tanh} \,\beta = 2p-1 ,
\label{nline}
\end{equation}
where $\beta\equiv 1/T$.  As a consequence of the
inequality~\cite{Nishimori-81}
\begin{equation}
|[\langle\sigma_x\sigma_y\rangle_T]|\le 
[|\langle\sigma_x\sigma_y\rangle_{T_N(p)}|]
\label{ineq}
\end{equation}
(the angular and the square brackets refer respectively to the thermal average
and to the quenched average over the bond couplings $\{J_{xy}\}$, while the
subscripts indicate the temperature of the thermal average), ferromagnetism
can only exist in the region $p\ge p^*$, and the system is maximally
magnetized along the $N$ line.  This implies that the PF boundary lies in the
region $p\ge p^*$.  At the MNP the transition line is predicted to be parallel
to the $T$ axis.\cite{LH-89} Then, it reaches the $T=0$ axis at $X_c=(0,p_c)$.
As a consequence of inequality (\ref{ineq}), $p_c$ must satisfy the inequality
\begin{equation}
p_c\ge p^*.
\label{pcs}
\end{equation}
At variance with the three-dimensional case, there is no evidence of a
finite-temperature glassy phase. Glassy behavior is only expected for $T=0$
and $p<p_c$ : the glassy phase at $T=0$ is unstable with respect to thermal
fluctuations.  In
Refs.~\onlinecite{ON-93,Kitatani-92,Nishimori-86,Nishimori-01} it was argued
that the PF transition line that connects the MNP to $X_c$ is only related to
the frustration distribution; hence, it should not depend on temperature and
should coincide with the line $p=p^*$, so that $p_c=p^*$.  This argument
provides a good approximation of the phase diagram below the MNP, although it
is not exact. Indeed, numerical analyses
\cite{PHP-06,AH-04,WHP-03,MC-02,BGP-98,KR-97} clearly support a reentrant
phase transition line with $p_c>p^*$. The difference is however quite small,
$p_c-p^*\approx 0.006$. The critical behavior along the transition line
connecting the MNP to the $T=0$ axis is an open issue.  Even though it
separates a paramagnetic phase from a ferromagnetic phase, it seems unlikely
that such transitions belong to the same universality class as the PF
transitions that occur on the line connecting the Ising point to the MNP. The
glassy transitions at $T=0$ and $p<p_c$ are expected to belong to the same
universality class as that of the bimodal model with $p=1/2$, see, e.g.,
Ref.~\onlinecite{JLMM-06} and references therein.  It is worth noting that the
point $X_c=(0,p_c)$ is a multicritical point: it is connected to three phases
and it is the intersection of two different transition lines, the PF line at
$T>0$ and the glassy line at $T=0$.  For $T=0$ the critical point $X_c$
separates a ferromagnetic phase from a glassy phase, while for $T > 0$ the
transition line separates a ferromagnetic from a paramagnetic phase.  The
behavior in a neighborhood of the multicritical point $X_c$ depends on the
nature of the transition.  If the PF transition and the glassy transition are
effectively decoupled, we expect a phase diagram like that reported in
Fig.~\ref{phdia}.  On the other hand, if the critical modes are coupled at
$X_c$, all transition lines should be tangent at the multicritical point; 
therefore the PF line should be tangent to the glassy transition line $T=0$.
Moreover, in this case the magnetic critical behavior at $T=0$ should differ
from that at $T>0$ along the transition line from the MNP to $X_c$.

Recently, Ref.~\onlinecite{NN-02} put forward an interesting conjecture
concerning the location of the MNP in a general class of models in
generic dimension.  In the case of the 2D $\pm J$ model it
predicts the MNP at
\begin{equation}
X_e\equiv (T_e=0.956729...,p_e=0.889972...).  
\label{nconje}
\end{equation}
The available numerical results show that Eq.~(\ref{nconje}) is a very good
approximation of the location of the MNP; for example, the transfer-matrix
calculations reported in Refs.~\onlinecite{MC-02},
\onlinecite{HPP-01} and \onlinecite{AQS-99} give $p^* = 0.8907(2)$,
0.8906(2), 0.8905(5), respectively.  Actually, since the small difference
$p^*-p_e \approx 0.0006$ corresponds at best to approximately three error
bars, these numerical works do not conclusively rule it
out.\cite{Nishimori-07} The conjecture has also been tested on hierarchical
lattices, where it has been found that it is not exact, although discrepancies
are numerically small~\cite{ONB-08,HB-05} also in this case.

In this paper we consider the square-lattice $\pm J$ model, determine
the location of the MNP, 
and study the critical behavior in its vicinity.  For this purpose,
we perform high-statistics Monte Carlo (MC) simulations along the $N$ line
close to the MNP. We consider lattices of size $L^2$ with $6\le L\le 64$.  
A detailed finite-size scaling
(FSS) analysis allows us to determine the location of the MNP quite precisely.
We obtain 
\begin{equation}
X_{\rm MNP}=[T^* = 0.9528(4), p^*=0.89081(7)].  
\end{equation}
We determine the renormalization-group (RG) dimensions $y_1$ and $y_2$ of the
relevant operators that control the RG flow close to the MNP.  We obtain $y_1
= 0.655(15)$ and $y_2 = 0.250(2)$, corresponding to the temperature and
crossover exponents $\nu\equiv 1/y_2=4.00(3)$ and 
$\phi\equiv y_1/y_2 = 2.62(6)$, respectively. 
Our results confirm that $X_e$ defined in Eq.~(\ref{nconje}) is 
a very good approximation of the MNP location:
indeed, $p^*-p_e=0.00084(7)$. However, they also show that the conjecture of
Ref.~\onlinecite{NN-02} leading to $X_e$ is not exact.

The paper is organized as follows.  In Sec.~\ref{sec2} we summarize
the theoretical results, focussing in particular on the FSS
behavior expected at the MNP.  In Sec.~\ref{sec3} we present
the FSS analysis of high-statistics MC simulations along the
$N$ line.  In Sec.~\ref{sec4} we summarize our results and draw our
conclusions.  In App.~\ref{notations} we report some notations.

\section{Finite-size scaling at the multicritical point}
\label{sec2}

In the absence of external fields, the critical behavior at the 
MNP is characterized by two relevant RG operators. The
singular part of the disorder-averaged free energy in a volume $L^d$ 
can be written as
\begin{equation}
F_{\rm sing}(T,p,L) = L^{-d} f(u_1 L^{y_1}, u_2 L^{y_2}, 
\{u_i L^{y_i}\}),\quad i\ge 3,
\label{freeen}
\end{equation} 
where $y_1>y_2>0$, $y_i<0$ for $i\ge 3$, $u_i$ are the corresponding scaling
fields, $u_1 = u_2 = 0$ at the MNP, and $d$ is the space dimension ($d=2$ in
the present case).  In the infinite-volume limit and neglecting 
scaling corrections due to irrelevant scaling fields,
we have
\begin{equation}
F_{\rm sing}(T,p) = |u_2|^{d/y_2} f_\pm (u_1 |u_2|^{-\phi}), 
\qquad \phi\equiv y_1/y_2>1,
\label{freeen2}
\end{equation} 
where the functions $f_\pm(x)$ apply to the parameter regions in which $\pm
u_2 > 0$. Close to the MNP, the transition lines correspond to constant values
of the product $u_1 |u_2|^{-\phi}$ and thus, since $\phi > 1$, they are
tangent to the line $u_1 = 0$.

The scaling fields $u_i$ are analytic functions of the model parameters $T$
and $p$.  Using symmetry arguments, Refs.~\onlinecite{LH-88,LH-89} showed that
one of the scaling axes is along the $N$ line, i.e., that the $N$ line is either
tangent to the line $u_1 = 0$ or to $u_2 = 0$. Since the $N$ line cannot be
tangent to the transition lines at the MNP and these lines are tangent to $u_1
= 0$, the first possibility is excluded. Thus, close to the MNP the $N$ line
corresponds to $u_2 = 0$.  Thus, we identify\cite{LH-88,LH-89}
\begin{equation}
u_2 =  {\rm tanh} \beta-2p+1.
\label{u2sf}
\end{equation} 
As for the scaling axis $u_1 = 0$, $\epsilon\equiv 6-d$ expansion calculations
predict it \cite{LH-89} to be parallel to the $T$ axis.  The extension of this
result to lower dimensions suggests
\begin{equation}
u_1=p-p^*.
\label{u1sf}
\end{equation} 
Note that, if Eq.~(\ref{u1sf}) holds, only the scaling field $u_2$ depends on
the temperature $T$. We may then identify $\nu=1/y_2$, and rewrite
Eq.~(\ref{freeen2}) as
\begin{equation}
F_{\rm sing}(T,p) = |t|^{2\nu} f_\pm ( g |t|^{-\phi}), 
\label{freeen3}
\end{equation} 
where $t\equiv (T-T^*)/T^*$, $g\equiv p-p^*$, and $\phi$ is the crossover
exponent.

These results give rise to the following predictions for the FSS behavior
around $T^*$, $p^*$. Let us consider a RG invariant quantity $R$, such as
$R_\xi\equiv \xi/L$, $U_4$, $U_{22}$, which are defined in the
App.~\ref{notations} and called phenomenological couplings.  In the FSS limit
$R$ obeys the scaling law
\begin{equation}
R = {\cal R}(u_1 L^{y_1}, u_2 L^{y_2}, \{u_i L^{y_i}\}),\quad i\ge 3.
\label{scalR}
\end{equation}
Neglecting the scaling corrections which vanish in the limit $L\to
\infty$, we expand in the neighborhood of the MNP: 
\begin{equation}
R = R^* + b_{11} u_1 L^{y_1} + b_{21} u_2 L^{y_2} + \ldots.
\label{scalR1}
\end{equation}
Along the $N$ line, the
scaling field $u_2$ vanishes, so that we can write
\begin{equation}
R_N = R^* + b_{11} u_1 L^{y_1} + \ldots,
\label{RGinvsca}
\end{equation}
where the subscript $N$ indicates that $R$ is restricted to the $N$ line. Let
us now consider the derivative of $R$ with respect to $\beta\equiv 1/T$.
Differentiating Eq.~(\ref{scalR1}), we obtain
\begin{equation}
R' = b_{11} u'_1 L^{y_1} + b_{21} u'_2 L^{y_2} + \cdots 
\label{RGinvdsca1}
\end{equation}
If Eq.~(\ref{u1sf}) holds, then $u'_1=0$, so that
\begin{equation}
R' = b_{21} u'_2 L^{y_2} + \cdots 
\label{RGinvdsca2}
\end{equation}
This result gives us a method to verify the conjecture of
Ref.~\onlinecite{LH-89}: once $y_1$ has been determined from the scaling
behavior of a RG invariant quantity $R$ close to the MNP, it is enough to
check the scaling behavior of $R'$. If $R'$ scales as $L^\zeta$ with
$\zeta<y_1$, the conjecture is confirmed and $\zeta$ provides an estimate of
$y_2$.  Along the $N$ line the magnetic susceptibility is expected to behave
as
\begin{equation}
\chi_N = e L^{2-\eta}\left( 1 + e_1 u_1 L^{y_1} + \cdots\right).
\label{RGinvchi}
\end{equation}
Let us mention that the general features of the MNP are expected to be
independent of $d$.  In three dimensions they have been accurately verified in
Refs.~\onlinecite{HPPV-07-mcpmj,Singh-91}.

\section{Monte Carlo results}
\label{sec3}

\subsection{Simulation details} \label{MNP-MC}

In the following we present a FSS analysis of high-statistics MC
data along the $N$ line defined by
\begin{equation}
\beta = \beta_N(p) \equiv  - {1\over 2} \ln\left({1-p\over p}\right).
\label{nline2}
\end{equation} 
We performed MC simulations on square lattices of linear size $L$ with
periodic boundary conditions, for several values of $L$,
$L=6,8,12,16,24,32,48,64$.  Most simulations were performed close to the MNP,
for values of $p$ in the range $0.8895\le p \le 0.8920$, which includes the
value $p_e = 0.889972\ldots$

We used a standard Metropolis algorithm up to $L=24$, while for $L\ge 32$ we
supplemented the updating method with the random-exchange technique\cite{raex}
(see also Sec. 3 in Ref.~\onlinecite{PTPV-06} for a discussion of the
random-exchange method in a disordered system).  In order to determine MC
estimates at $p$ and $\beta = \beta_N(p)$, we considered $N_T$ systems at the
same value of $p$ and at inverse temperatures $\beta_{\rm min} \equiv
\beta_1$, \ldots, $\beta_{N_T} = \beta_N(p)$. The chosen values of $\beta$
were equally spaced, i.e.  $\beta_{i+1} - \beta_i = \Delta\beta$, with a
constant $\Delta\beta$ (typically, $\Delta\beta \approx 0.06,0.04,0.03$ for
$L=32,48$, and 64).  The spacing $\Delta\beta$ was chosen such that the
acceptance probability was significantly larger than zero, while $\beta_{\rm
  min}$ was chosen to have a sufficiently fast thermalization at $\beta =
\beta_{\rm min}$.  The elementary unit of the algorithm consisted in $N_{\rm
  ex}$ Metropolis sweeps for each configuration followed by an exchange move.
We considered all pairs of configurations corresponding to nearby temperatures
and proposed a temperature exchange with acceptance probability
\begin{equation}
  {\cal P} = \exp \{(\beta_i - \beta_{i+i})(E_i-E_{i+1})\},
\end{equation}
where $E_i$ is the energy of the system at inverse temperature $\beta_i$. In
our MC runs we chose $N_{\rm ex} = 20$.  In our simulations we used multispin
coding (details can be found in Ref.~\onlinecite{HPPV-07-pmj}).

In Table~\ref{table_parallel} we report the parameters of our simulations
performed with the random-exchange algorithm: here $N_{\rm run}$ is the number
of Metropolis sweeps per sample and temperature, while $N_{\rm therm}$ is the
corresponding number of Metropolis sweeps discarded for thermalization. We
also report the range of the exchange probability, which depends on the
temperatures $\beta_i$ and $\beta_{i+1}$ considered.

\begin{table}
\squeezetable
\caption{
Parameters of our random-exchange MC runs. $N_{\rm run}$ is the
    number of Metropolis sweeps per configuration and sample, 
    $N_{\rm therm}$ is the corresponding number of 
    Metropolis sweeps discarded for thermalization.}
\label{table_parallel}
\begin{ruledtabular}
  \begin{tabular}{cllccrr}
\multicolumn{1}{c}{$L$}&
\multicolumn{1}{c}{$p$}&
\multicolumn{1}{c}{$\beta_{\rm min}$}&
\multicolumn{1}{c}{$N_T$}&
\multicolumn{1}{c}{acc. range}&
\multicolumn{1}{c}{$N_{\rm run}/10^3$}&
\multicolumn{1}{c}{$N_{\rm therm}/10^3$}
\\
\colrule
    $32$ & $0.8895$   & $0.3228$   &  $13$ & $4\%-56\%$ & 240 & $48$ \\
    $32$ & $0.889972$ & $0.3252$   &  $13$ & $4\%-56\%$ & 240 & $48$ \\
    $32$ & $0.8905$   & $0.3279$   &  $13$ & $4\%-56\%$ & 240 & $48$ \\
    $32$ & $0.8910$   & $0.3305$   &  $13$ & $4\%-57\%$ & 240 & $72$ \\
    $32$ & $0.8915$   & $0.3331$   &  $13$ & $4\%-57\%$ & 240 & $48$ \\
    $48$ & $0.889972$ & $0.285228$ &  $20$ & $4\%-57\%$ & 400 & $80$ \\
    $48$ & $0.8905$   & $0.2879$   &  $20$ & $4\%-57\%$ & 400 & $80$ \\
    $48$ & $0.8910$   & $0.2905$   &  $20$ & $4\%-57\%$ & 400 & $80$ \\
    $48$ & $0.8915$   & $0.3310$   &  $25$ & $13\%-66\%$& 320 & $64$ \\
    $64$ & $0.889972$ & $0.265228$ &  $27$ & $4\%-57\%$ & 900 &$180$ \\
    $64$ & $0.8906$   & $0.268442$ &  $27$ & $4\%-57\%$ & 600 &$180$ \\
    $64$ & $0.8909$   & $0.2700$   &  $27$ & $4\%-58\%$ & 600 &$300$ \\
    $64$ & $0.8909$   & $0.2700$   &  $27$ & $4\%-58\%$ &1200 &$240$ \\
    $64$ & $0.8912$   & $0.271529$ &  $27$ & $4\%-58\%$ & 600 &$240$ \\
\end{tabular}
\end{ruledtabular}
\end{table}

For every disorder sample we performed a MC run of $N_{\rm run}$ Metropolis
sweeps, collecting $N_{\rm meas}$ measures of the quantities defined in
App.~\ref{notations}.  We used $N_{\rm meas}=400$ for $L\le 24$ and $N_{\rm
  meas} = 100$ for $L \ge 32$.  In order to obtain equilibrated data, we
discarded a fraction of the measures which is determined by using the
following procedure. We divided the measures into $N_B$ parts (tipically
$N_B=10$, $20$) of length $l=N_{\rm meas}/N_B$. Then, we considered the
disorder-averaged susceptibilities
\begin{equation}
  \chi_b(t) = \left[ \frac{1}{l} \sum_{i = t l}^{(t+1)l - 1} \chi(i) \right], 
\qquad t=0 \ldots N_B-1.
\end{equation}
Starting from random infinite-temperature spin configurations, $\chi_b(t)$
increases with $t$.  When $t$ is sufficiently large, $\chi_b(t)$ becomes
constant within error bars, thus signalling that thermalization has been
reached.  We considered the susceptibility because it is expected to be
particularly sensitive to thermalization.  Whenever one determines disorder
averages of functions of thermal averages one should perform a bias
correction; for this purpose we used the results of Ref.~\onlinecite{HPPV-07}.

MC results are reported in Tables~\ref{tabn1} and \ref{tabn2}.  To obtain
small statistical errors, we generated a large number of samples $N_s$:
$N_s=10^6$ in all cases, except for the run with $L=32$ and $p=0.891$, where
$N_s=4\times 10^6$. An important check of our simulations is given by the
comparison with the exact behavior of the energy density along the $N$
line,\cite{Nishimori-81}
\begin{equation}
E_N(p) = {1\over V} [ \langle {\cal H} \rangle_{T_N(p)} ] = 2-4p.
\label{energy}
\end{equation}
All runs give estimates of $E_N(p)$ which are consistent with 
Eq.~(\ref{energy}).
For example, we obtain $E_N(p)/(2-4p)=1.00001(1),\,1.00000(1)$
for $L=32,\,p=0.891$ and $L=64,\,p=0.8909$, respectively.

\begin{table}
\squeezetable
\caption{
MC data for $L=6,8,12,16$ along the $N$ line.
For all runs the number of samples is $N_s=10^6$.
}
\label{tabn1}
\begin{ruledtabular}
\begin{tabular}{rlllllll}
\multicolumn{1}{c}{$L$}&
\multicolumn{1}{c}{$p$}&
\multicolumn{1}{c}{$R_\xi$}&
\multicolumn{1}{c}{$U_4$}&
\multicolumn{1}{c}{$U_{22}$}&
\multicolumn{1}{c}{$U_d$}&
\multicolumn{1}{c}{$\chi$}&
\multicolumn{1}{c}{$R'_\xi$}
\\
\colrule
6 & 0.8895   &0.9806(8) &1.1316(2) &0.08302(17) &1.04856(7) &26.172(8) & 6.532(6) \\
  & 0.889972 &0.9886(8) &1.1295(2) &0.08176(17) &1.04776(7) &26.270(8) & 6.456(6) \\
  & 0.8905   &0.9979(8) &1.1272(2) &0.08032(17) &1.04687(6) &26.382(8) & 6.371(6) \\ 
  & 0.891    &1.0068(8) &1.1250(2) &0.07899(17) &1.04604(6) &26.486(8) & 6.290(6) \\
  & 0.8915   &1.0158(8) &1.1229(2) &0.07765(17) &1.04521(6) &26.590(8) & 6.210(6) \\
  & 0.892    &1.0251(8) &1.1207(2) &0.07632(16) &1.04438(6) &26.695(8) & 6.129(6) \\
\colrule
8 & 0.8895   &0.9741(7) &1.1327(2) &0.08455(17) &1.04810(6) &44.150(13) & 13.471(11)\\
  & 0.889972 &0.9837(7) &1.1301(2) &0.08301(16) &1.04713(6) &44.366(13) & 13.307(11) \\
  & 0.8905   &0.9945(7) &1.1274(2) &0.08132(16) &1.04609(6) &44.604(13) & 13.126(11) \\
  & 0.891    &1.0049(7) &1.1249(2) &0.07973(16) &1.04512(6) &44.829(13) & 12.953(11) \\
  & 0.8915   &1.0156(7) &1.1223(2) &0.07811(16) &1.04418(6) &45.054(12) & 12.787(11) \\
  & 0.892    &1.0265(8) &1.1198(2) &0.07654(16) &1.04323(6) &45.278(12) & 12.615(11) \\
\colrule
12& 0.8895   &0.9630(7) &1.1350(2) &0.08672(17) &1.04831(6) &91.98(3) & 36.01(3) \\
  & 0.889972 &0.9754(7) &1.1317(2) &0.08464(17) &1.04707(6) &92.61(3) & 35.56(3) \\
  & 0.8905   &0.9893(7) &1.1281(2) &0.08239(17) &1.04572(6) &93.31(3) & 35.02(2) \\
  & 0.891    &1.0030(7) &1.1247(2) &0.08028(16) &1.04444(6) &93.97(3) & 34.52(2) \\
  & 0.8915   &1.0167(8) &1.1214(2) &0.07821(16) &1.04322(6) &94.63(3) & 34.04(2) \\
  & 0.892    &1.0309(8) &1.1182(2) &0.07618(16) &1.04202(6) &95.29(3) & 33.56(2) \\
\colrule
16& 0.8895   &0.9554(7) &1.1373(2) &0.08840(16) &1.04891(7) &154.69(5) & 71.09(4) \\
  & 0.889972 &0.9701(7) &1.1333(2) &0.08587(16) &1.04740(6) &156.02(5) & 70.15(4) \\
  & 0.8905   &0.9870(7) &1.1289(2) &0.08311(16) &1.04577(6) &157.52(5) & 69.11(4) \\
  & 0.891    &1.0033(7) &1.1248(2) &0.08051(16) &1.04428(6) &158.92(5) & 68.10(4) \\
  & 0.8915   &1.0199(7) &1.1208(2) &0.07804(15) &1.04280(6) &160.31(4) & 67.06(4) \\
  & 0.892    &1.0367(8) &1.1170(2) &0.07562(15) &1.04140(6) &161.68(4) & 66.03(4) \\
\end{tabular}
\end{ruledtabular}
\end{table}

\begin{table}
\squeezetable
\caption{
  MC data for $L=24,32,48,64$ along the $N$ line.
  The number of samples is $N_s=10^6$,
  except for the run with $L=32$ and $p=0.891$. In this case
  $N_s=4\times 10^6$.
}
\label{tabn2}
\begin{ruledtabular}
\begin{tabular}{rlllllll}
\multicolumn{1}{c}{$L$}&
\multicolumn{1}{c}{$p$}&
\multicolumn{1}{c}{$R_\xi$}&
\multicolumn{1}{c}{$U_4$}&
\multicolumn{1}{c}{$U_{22}$}&
\multicolumn{1}{c}{$U_{d}$}&
\multicolumn{1}{c}{$\chi$}&
\multicolumn{1}{c}{$R'_\xi$}
\\
\colrule
24 & 0.8895   &0.9423(7) &1.1411(2) &0.09071(18) &1.05037(7) &321.10(10) &182.16(12) \\
   & 0.889972 &0.9615(7) &1.1356(2) &0.08730(17) &1.04833(7) &324.95(9)  &179.74(11) \\
   & 0.8905   &0.9831(7) &1.1299(2) &0.08373(16) &1.04613(6) &329.20(9)  &176.77(11) \\
   & 0.891    &1.0042(7) &1.1247(2) &0.08054(16) &1.04411(6) &333.15(9)  &173.76(11) \\
   & 0.8915   &1.0265(8) &1.1194(2) &0.07724(15) &1.04217(6) &337.14(10) &170.92(11) \\
   & 0.892    &1.0486(8) &1.1145(2) &0.07412(15) &1.04038(6) &341.06(9)  &168.12(10) \\
\colrule
32 & 0.8895   &0.9319(6) &1.1443(2)  &0.09279(18) &1.05152(7) &538.29(16) &351.4(3) \\
   & 0.889972 &0.9533(7) &1.1380(2)  &0.08878(18) &1.04922(7) &546.13(17) &346.3(3) \\  
   & 0.8905   &0.9808(7) &1.1306(2)  &0.08423(17) &1.04640(7) &555.14(17) &334.0(3) \\
   & 0.891    &1.0058(4) &1.12439(10)&0.080358(8) &1.04404(3) &563.43(8)  &334.02(14) \\
   & 0.8915   &1.0315(8) &1.1184(2)  &0.07651(16) &1.04189(6) &571.52(16) &328.6(3) \\
   & 0.892    &1.0587(8) &1.11249(18)&0.07292(15) &1.03958(5) &579.70(16) &322.2(2) \\
\colrule
48 & 0.889972 &0.9430(7) &1.1413(2)  &0.09079(17) &1.05050(7) &1134.1(3) &864.7(8) \\
   & 0.8905   &0.9758(7) &1.1321(2)  &0.08498(17) &1.04715(7) &1158.4(3) &847.8(8) \\
   & 0.891    &1.0088(7) &1.1237(2)  &0.07984(16) &1.04390(6) &1181.8(3) &830.2(7) \\
   & 0.8915   &1.0438(8) &1.11581(18)&0.07487(15) &1.04094(6) &1204.9(3) &814.2(7) \\
\colrule
64 & 0.889972 &0.9311(6) &1.1449(2) &0.09308(17) &1.05186(7) &1900.0(6) &1641.3(1.6) \\
   & 0.8906   &0.9792(7) &1.1313(2) &0.08460(17) &1.04670(6) &1961.0(6) &1596.9(1.6) \\
   & 0.8909   &1.0037(7) &1.1252(2) &0.08072(17) &1.04444(6) &1990.3(6) &1580.2(1.5)\\
   & 0.8912   &1.0286(8) &1.1194(2) &0.07711(16) &1.04226(6) &2019.1(6) &1560.2(1.6) \\
\end{tabular}
\end{ruledtabular}
\end{table}

\begin{figure*}[tb]
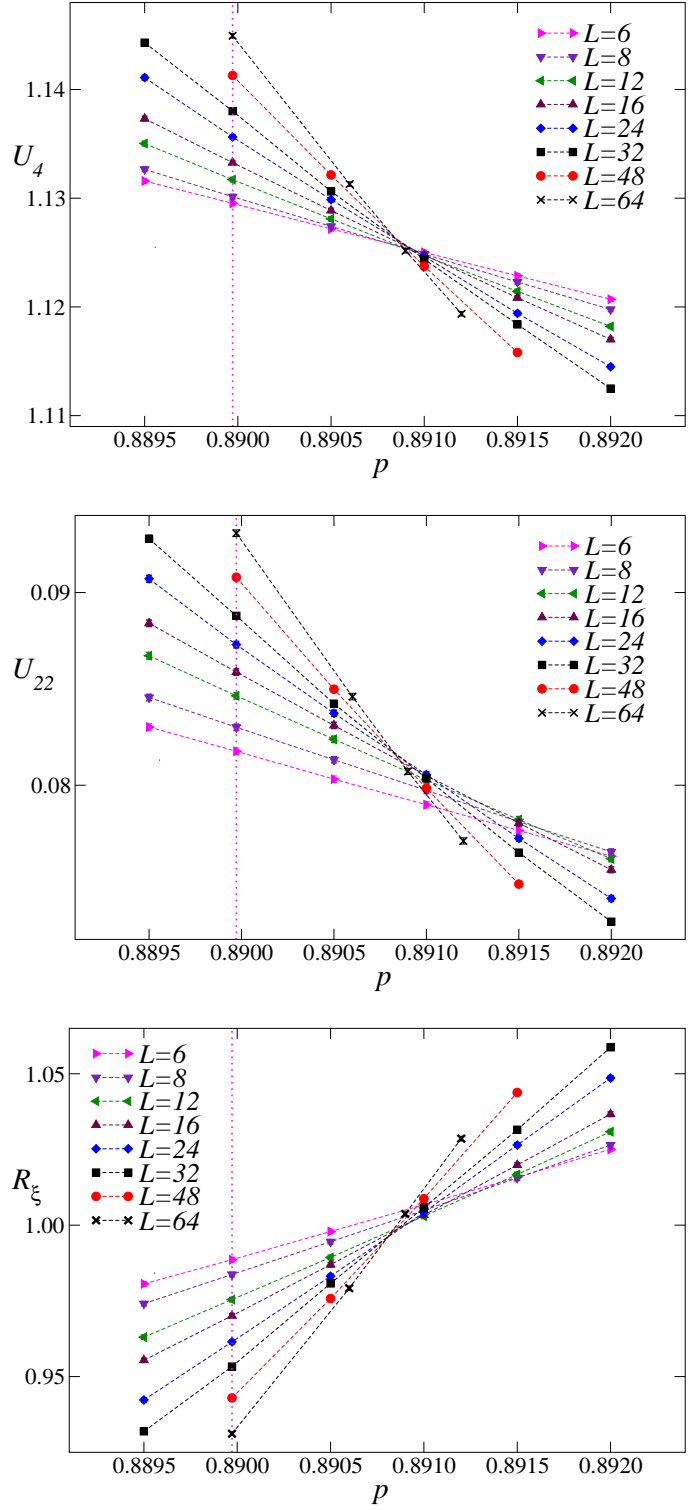

\centerline{\psfig{width=9truecm,angle=0,file=u4.eps}}
\vspace{4mm}
\centerline{\psfig{width=9truecm,angle=0,file=u22.eps}}
\vspace{4mm}
\centerline{\psfig{width=9truecm,angle=0,file=rxi.eps}}
\caption{(Color online)
  MC data of $U_4$, $U_{22}$ and $R_\xi\equiv \xi/L$ vs $p$.  
  The dashed lines connecting
  the data at given $L$ are drawn to guide the eye.  The dotted vertical line
  corresponds to $p=p_e=0.889972$. 
}
\label{rgifig}
\end{figure*}

\subsection{Results} \label{secres}

MC estimates of the RG invariant quantities $R_\xi$, $U_4$, and $U_{22}$ along
the $N$ line are shown in Fig.~\ref{rgifig}.  There is clearly a crossing
point at $p \approx 0.891$.  The raw data already indicate that
$p^*>p_e=0.889972..$, where $p_e$ is the value conjectured in
Ref.~\protect\onlinecite{NN-02}. Their difference can hardly be explained in
terms of scaling corrections. Indeed, the crossing point $p_{\rm
  cross}(L,\kappa)$ of the data corresponding to lattice sizes $L$ and $\kappa
L$ scales for $L\to\infty$ as
\begin{equation}
p_{\rm cross}(L,\kappa) - p^* \sim L^{-y_1 - \omega},
\label{convp}
\end{equation}
where $\omega>0$ is the exponent associated with the leading irrelevant
operator.  Since, as we shall see, $y_1\approx 0.6$, the approach is
reasonably fast, so that our data, that correspond to lattice sizes between 6
and 64, should be able to detect a drift due to scaling corrections. The very
good stability of the results excludes a delayed approach to $p_e$.

In order to estimate precisely $p^*$, $T^*$, and $y_1$, we perform a FSS
analysis of the phenomenological couplings $R_\xi\equiv \xi/L$, $U_4$,
$U_{22}$, and $U_d$, which are defined in App.~\ref{notations} and are
generically denoted by $R$.  Since we vary $p$ and $\beta$ along the $N$ line,
close to the MNP we expect
\begin{equation}
R = f_R[ (p - p^*) L^{y_1}],
\end{equation}
with $f_R(0) = R^*$. This functional form relies on the property that $u_2 = 0$
along the $N$ line.  Since our data are sufficiently close to the MNP, the
product $(p - p^*) L^{y_1}$ is small. We can thus expand $f_R(x)$ in powers of
$x$. Thus, we fit the numerical data to
\begin{equation}
   R = R^* + \sum_{n=1}^{n_{\rm max}} a_n (p - p^*)^n L^{ny_1}, 
\label{fitR-1}
\end{equation}
keeping $R^*$, the coefficients $\{a_n\}$, $p^*$, and $y_1$ as free
parameters. Here we neglect scaling corrections.  To monitor their
role, we repeat the fits several times, each time only including data
satisfying $L\ge L_{\rm min}$.  Fits with $n_{\rm max} = 1$ have a large
$\chi^2/{\rm DOF}$ (DOF is the number of degrees of freedom of the fit),
indicating that the range of values of $p$ we are considering is too large to
allow for a linear approximation of the scaling function $f_R(x)$.  Fits with
$n_{\rm max} = 2$ have instead a good $\chi^2$ for $L_{\rm min} \ge 6$
($U_4$), 12 ($R_\xi$), and 16 ($U_{22}$ and $U_d$).  We also perform fits
with $n_{\rm max} = 3$, but we do not observe significant differences: for
$U_d$ and $L_{\rm min} = 6,12$ we obtain $\chi^2/{\rm DOF} = 2148/39,34/27$
with $n_{\rm max} = 2$, and $2147/38,34/26$ for $n_{\rm max} = 3$. Clearly, a
parabolic approximation is fully adequate. Beside fitting separately each
observable, we also perform combined fits of three different phenomenological
couplings.  The results are reported in Table \ref{fit-R-nocorr}. In the case
of $U_{22}$, $U_4$, and $R_\xi$ all estimates of $p^*$ show a systematic
downward trend with $L_{\rm min}$, with $0.89080\lesssim p^*\lesssim 0.89083$
for $L_{\rm min} = 32$. Fits of $U_d$ (note that this quantity is
statistically more precise than the other ones, see the Tables~\ref{tabn1} and
\ref{tabn2}, which explains the somewhat
larger $\chi^2/{\rm DOF}$ of the fits) show instead a different behavior and
suggest a somewhat larger value of $p^*$, $p^*\approx 0.89087$.  Similar
trends are observed in the estimates of $y_1$, which in most of the cases
increases with $L_{\rm min}$ and varies essentially in the range $0.65
\lesssim y_1 \lesssim 0.67$ with a statistical error of $\pm 0.01$-0.02.

\begin{table}
\caption{
Estimates of $p^*$ and $y_1$ obtained by performing a fit to 
Eq.~(\ref{fitR-1}) with $n_{\rm max} = 2$. 
DOF is the number of 
degrees of freedom in the fit and $L_{\rm min}$ is the minimum lattice size 
included in the fit.
}
\label{fit-R-nocorr}
\begin{ruledtabular}
\begin{tabular}{cccll}
    & $L_{\rm min}$ & $\chi^2/{\rm DOF}$ &        $p^*$    &     $y_1$ \\
   \hline 
  $U_d$  &  12      &        34/27     &      0.890860(7)  &  0.658(7) \\
         &  16      &        19/21     &      0.890877(8)  &  0.656(9) \\
         &  24      &        17/15     &      0.890881(11) &  0.647(11)\\
         &  32      &         14/9     &      0.890871(14) &  0.664(16)\\
\hline
$U_{22}$ &  12      &        41/27     &      0.890895(12) &  0.639(11)\\
         &  16      &        11/21     &      0.890857(13) &  0.647(13)\\
         &  24      &         6/15     &      0.890831(17) &  0.663(18)\\
         &  32      &          3/9     &      0.890813(21) &  0.672(25)\\
\hline
$U_4$    &  12      &        19/27     &      0.890882(9)  &  0.647(9) \\
         &  16      &         9/21     &      0.890865(10) &  0.650(10)\\
         &  24      &         6/15     &      0.890850(13) &  0.656(14)\\
         &  32      &          3/9     &      0.890834(17) &  0.669(19)\\
\hline
$R_\xi$  &  12      &        21/27     &      0.890835(8)  &  0.647(8) \\
         &  16      &        12/21     &      0.890820(9)  &  0.650(9) \\
         &  24      &         7/15     &      0.890805(12) &  0.653(12)\\
         &  32      &          5/9     &      0.890795(15) &  0.664(17)\\
\hline
$R_\xi$,$U_4$,$U_{22}$
         &  12      &        107/85    &      0.890864(5)  &   0.646(5)\\
         &  16      &         44/67    &      0.890844(6)  &   0.650(6)\\
         &  24      &         27/49    &      0.890826(8)  &   0.656(8)\\
         &  32      &         14/31    &      0.890812(10) &   0.668(11)\\
\hline
$R_\xi$,$U_4$,$U_{d}$
         &  12      &         92/85    &      0.890858(4)  &   0.651(4) \\
         &  16      &         64/67    &      0.890855(5)  &   0.653(5) \\
         &  24      &         54/49    &      0.890847(7)  &   0.652(7) \\
         &  32      &         37/31    &      0.890836(9)  &   0.665(10)\\
\end{tabular}
\end{ruledtabular}
\end{table}

These tiny discrepancies indicate that scaling corrections are not negligible
if compared with our small statistical errors. In order to estimate their
quantitative role, we also perform fits in which 
scaling corrections are taken into account. Thus, we fit the MC data  to
\begin{equation}
   R = R^* + \sum_{n=1}^{n_{\rm max}} a_n (p-p^*)^n L^{ny_1} + 
       L^{-\omega} \sum_{k=0}^{k_{\rm max}} b_k (p-p^*)^k L^{ky_1}.
\label{fitR-2}
\end{equation}
Results for $k_{\rm max} = 0$ and 1 have both a good $\chi^2/{\rm DOF}$, even
for $L_{\rm min} = 6$. In the following we present results corresponding to
$k_{\rm max} = 1$, since this choice allows us to take into account the
scaling corrections that affect the determination of both $p^*$ and $y_1$. The
correction-to-scaling exponent $\omega$ is not known and thus we keep it
as a free parameter.  Our results are reported in Table~\ref{fit-R-corr}.
Because of the large number of parameters this fit gives stable results only
for $L_{\rm min} = 6,8$ (for $U_4$ this is not even the case). For larger
values of $L_{\rm min}$ errors are so large to make the results meaningless.
The results are fully consistent. First of all, they predict $\omega \gtrsim
1$.  Thus, corrections to scaling decay reasonably fast, indicating that the
systematic error should be reasonably estimated by considering data in our
range $6\le L \le 64$. Second, fits that involve $U_d$ give
estimates of $\omega$ that are significantly larger. This is consistent with
the results reported in Table~\ref{fit-R-nocorr}: fits involving $U_d$ show a
small dependence on $L_{\rm min}$, suggesting that $U_d$ is less affected by
scaling corrections. The estimates of $p^*$ obtained in fits of $\xi/L$ and
$U_{22}$ show a trend that is opposite to that observed in fits without
corrections, indicating that the correct value for $p^*$ belongs to the range
of values that occur in the two types of fits: values smaller than 0.89070 are
not consistent with our data.  To quote a final result, let us note that the
fits with $L_{\rm min} = 8$ reported in Table~\ref{fit-R-corr} give
(including the statistical error) $0.89074\lesssim p^* \lesssim 0.89089$. A
conservative estimate is therefore
\begin{equation}
p^* = 0.89081(7).
\label{p-est}
\end{equation}
This result is fully consistent with those obtained in the fits without 
scaling corrections. Using Eq.~(\ref{nline}) we obtain 
\begin{equation}
  \beta^* = 1.0495(4), \qquad T^* = 0.9528(4).
\end{equation}
Note that the conjectured value\cite{NN-02} $p_e = 0.889972\ldots$ 
is excluded, the difference $p^* - p_e=0.00084(7)$ 
corresponding to 12 error bars.

\begin{table}
\caption{
Estimates of $p^*$, $y_1$, and $\omega$ obtained by performing a fit to 
Eq.~(\ref{fitR-2}) with $n_{\rm max} = 2$ and $k_{\rm max} = 1$.
DOF is the number of 
degrees of freedom in the fit and $L_{\rm min}$ is the minimum lattice size 
included in the fit.
}
\label{fit-R-corr}
\begin{ruledtabular}
\begin{tabular}{ccclll}
& $L_{\rm min}$ & $\chi^2/{\rm DOF}$ &  $p^*$ &  $y_1$   & $\omega$ \\
\hline
$\xi/L$   &  6  &  9.9/36 &  0.89077(2) &    0.663(16) &  1.18(31) \\
          &  8  &  7.3/30 &  0.89079(2) &    0.654(13) &  1.95(68) \\
$U_{22}$  &  6  &  7.6/36 &  0.89077(3) &    0.663(21) &  1.23(25) \\
          &  8  &  7.1/30 &  0.89078(4) &    0.660(23) &  1.45(46) \\
$U_{d}$   &  6  &   31/36 &  0.89090(1) &    0.655(8)  &  2.67(16) \\
          &  8  &   20/30 &  0.89088(1) &    0.654(8)  &  4.15(68) \\
$\xi/L$,$U_4$,$U_{22}$
          &  6  &  77/114 &  0.89082(1) &    0.657(8)  &  1.59(17) \\
          &  8  &   47/96 &  0.89081(1) &    0.657(10) &  1.64(32) \\
$\xi/L$,$U_4$,$U_{d}$
          &  6  & 128/114 &  0.890868(5)&    0.652(5)  &  3.04(12) \\
          &  8  &   81/96 &  0.890860(5)&    0.651(5)  &  5.25(71) \\
\end{tabular}
\end{ruledtabular}
\end{table}

Let us finally estimate $y_1$. Fits with scaling corrections give results
that decrease with $L_{\rm min}$, while fits without scaling corrections
give estimates that have the opposite trend. Comparing all results, we 
infer $0.64\lesssim y_1 \lesssim 0.67$, so that we arrive at the final 
estimate
\begin{equation}
y_1 = 0.655(15).
\label{y1-est}
\end{equation}
The fits that we have reported also allow us to estimate the critical-point 
value $R^*$ of the phenomenological couplings. We obtain:
\begin{eqnarray}
R_\xi^* &=& 0.996(2), \\
U_4^* &=& 1.1264(6), \\
U_{22}^* &=& 0.0817(5), \label{u22est}\\
U_{d}^* &=& 1.0447(3). 
\end{eqnarray}
Of course, the estimate of $U_d^*$ is consistent with the relation 
$U_d^*=U_4^*-U_{22}^*$.
Note that these results are not very much different from those of the pure 2D
Ising values that apply along the PF line from the pure
Ising point at $p=1$ to the MNP, which are~\cite{SS-00} $R_\xi^* =
0.9050488292(4)$, $U_4^* = U_d^* = 1.167923(5)$, $U_{22}^* = 0$. In
particular, the estimate (\ref{u22est}) of $U_{22}^*$ is quite small,
indicating that the violations of self-averaging are small.

Let us now consider the derivatives $R'$ of the phenomenological couplings.
Close to the MNP, $R'$ is expected to behave as 
\begin{equation}
   R' = L^\zeta f_{R'}[ (p-p^*) L^{y_1} ],
\end{equation}
where we have used the fact that along the $N$ line $u_2 = 0$. 
If Eq.~(\ref{u1sf}) holds,\cite{LH-89} we have additionally $\zeta = y_2$. 
To determine $\zeta$ we perform analyses analogous to those used before 
to determine $p^*$ and $y_1$. We expand $f_{R'}(x)$ in powers of $x$
and thus fit $R'$ to 
\begin{equation}
   \ln R' = \zeta \ln L + \sum_{n=0}^{n_{\rm max}} a_n (p-p^*)^n L^{ny_1}. 
\label{fitQ-1}
\end{equation}
We always fix $y_1$ to the value (\ref{y1-est}) and $p^*$ to the value
(\ref{p-est}), including in the final error the variation of $y_1$ and $p^*$
within one error bar.  As in the fits of $R$, we check the role of $n_{\rm
  max}$.  A significant improvement in the quality of the fit is observed by
changing $n_{\rm max}$ from 1 to 2, while no significant change is obtained by
increasing it to 3. Therefore, we fix $n_{\rm max}=2$.

\begin{table}
\caption{
Estimates of the exponent $\zeta = y_2$ and $\zeta = \eta$ obtained by 
performing a fit to Eq.~(\ref{fitQ-1}) with $n_{\rm max} = 2$.
}
\label{table-zeta}
\begin{ruledtabular}
\begin{tabular}{crcl}
   & $L_{\rm min}$ &  $\chi^2/{\rm DOF}$ &  $\zeta$  \\
\hline
$R_\xi'$   
  &   8  &    20/29   &  0.252(2) \\
  &  12  &    16/24   &  0.251(2) \\
  &  16  &    13/19   &  0.250(2) \\
  &  24  &    12/14   &  0.249(3) \\
$U_4'$
  &    8 &    19/29   &  0.250(1) \\
  &   12 &    17/24   &  0.249(1) \\
  &   16 &    14/19   &  0.249(1) \\
  &   24 &    13/14   &  0.250(2) \\
$Z$   
  &    8 &   147/29 &  0.173(3) \\
  &   12 &    38/24 &  0.175(3) \\
  &   16 &    21/19 &  0.176(4) \\
  &   24 &    11/14 &  0.178(5)\\
  &   32 &     6/9  &  0.179(5)\\
\end{tabular}
\end{ruledtabular}
\end{table}

The results are reported in Table~\ref{table-zeta}. They are very stable and
show a very small dependence on $L_{\rm min}$, of the order of the statistical
error. As a final result we quote $\zeta = 0.250(2)$.  This result is
significantly smaller than $y_1$ and thus confirms the multicritical nature
of the MNP and the arguments reported
in Sec.~\ref{sec2}. Therefore $\zeta$ should be identified with
$y_2$, so that 
\begin{equation}
y_2 = 0.250(2), \qquad \nu \equiv {1\over y_2} = 4.00(3).
\end{equation}
The corresponding crossover exponent, cf. Eq.~(\ref{freeen2}), is 
\begin{equation}
\phi \equiv {y_1\over y_2} = 2.62(6).
\end{equation}
The same analysis used to estimate $y_2$ can be employed to determine $\eta$.
Instead of $\chi$, we consider
\begin{equation}
Z \equiv \chi/\xi^2\sim L^{-\eta}
\label{zedef}
\end{equation}
which has smaller statistical errors.  We fit $Z$ to
\begin{equation}
   \ln Z = -\eta \ln L + \sum_{n=0}^{n_{\rm max}} a_n (p-p^*)^n L^{ny_1}. 
\end{equation}
As before, we fix $y_1$ and $p^*$, set $n_{\rm max} = 2$, and repeat the fit
several times, each time considering only data satisfying $L\ge L_{\rm min}$.
The final results are reported in Table ~\ref{table-zeta}. A good $\chi^2$ is
obtained only for $L_{\rm min} \ge 16$.  The corresponding fits give $\eta
\approx 0.175$-0.180 with a slight upward trend. This effect may be real and
due to scaling corrections.  Therefore, we also fit $\ln Z$ to
\begin{equation}
   \ln Z = -\eta \ln L + \sum_{n=0}^{n_{\rm max}} a_n (p-p^*)^n L^{ny_1} + 
     L^{-\omega} \sum_{k=0}^{k_{\rm max}} a_k (p-p^*)^k L^{ky_1},
\label{fitQ-2}
\end{equation}
fixing $y_1$ and $p^*$, and keeping $\omega$ as a free parameter.
For $n_{\rm max} = 2$ and $k_{\rm max} = 1$, we obtain a good $\chi^2$ 
for any $L_{\rm min} \ge 6$. The corresponding estimates of $\eta$ are:
$\eta = 0.182(10)$ ($L_{\rm min} = 6$) and 
$\eta = 0.181(10)$ ($L_{\rm min} = 8$). The central estimates are 
quite close to those obtained in fits without scaling corrections, indicating 
that scaling corrections are small. 
We take as our final estimate
\begin{equation}
\eta =  0.180(5),
\end{equation}
which includes all results without scaling corrections and is consistent 
with the fits in which scaling corrections are taken into account.

\section{Conclusions} \label{sec4}

In this paper we have investigated the critical behavior 
of the square-lattice $\pm J$ Ising model close to the  MNP.
Our main results are the following:
\begin{itemize}
\item[(i)] We have obtained an accurate estimate of the location of the MNP:
  $p^* = 0.89081(7)$, $T^* = 0.9528(4)$.  The conjectured value
  $p_e=0.889972..$, put forward in Ref.~\onlinecite{NN-02},
  is a very good approximation, but it is not exact: 
  $p^*-p_e=0.00084(7)$.
\item[(ii)] 
We have computed the RG dimensions of the relevant operators at the MNP,
obtaining $y_1 = 0.655(15)$ and $y_2 = 0.250(2)$. 
It is tempting to conjecture that $y_2 = 1/4$ exactly. Note also
that $y_1$ is consistent with 2/3, though in this case the precision
of the result is not good enough to put this conjecture on firm grounds.
The above estimates of the RG dimensions give
$\nu \equiv  1/y_2 = 4.00(3)$ and $\phi \equiv y_1/y_2 = 2.62(6)$.
\item[(iii)]
We have computed the critical exponent $\eta$ that controls the critical
behavior of the magnetic correlations, obtaining 
$\eta = 0.180(5)$.
\end{itemize}

Our estimate of $p^*$ is significantly more precise than those obtained in
previous works.  By using transfer-matrix methods
Refs.~\onlinecite{MC-02,HPP-01,AQS-99,ON-87} obtained
$p^*=0.8907(2),\,0.8906(2),\,0.8905(5),\,0.889(2)$, respectively.  We also
mention the results $p^*=0.8872(8)$ obtained by means of an off-equilibrium MC
simulation,\cite{OI-98} and $p^*=0.886(3)$ from the analysis of
high-temperature expansions.\cite{SA-96} Concerning the critical exponents at
the MNP, we mention the square-lattice results $y_1=0.676(14)$, 0.667(13),
0.752(17), 0.75(7), 0.76(5)
respectively from Refs.~\onlinecite{PHP-06},%
\onlinecite{MC-02},\onlinecite{HPP-01},\onlinecite{CF-97},\onlinecite{SA-96}.
The estimate~\cite{Queiroz-06} $y_1=0.671(9)$ has been obtained on the
triangular and honeycomb lattices.  Moreover, we mention the
result~\cite{PHP-06} $y_1=0.658(13)$ obtained from a model with Gaussian
distributed couplings.  The most recent results are clearly consistent with
our estimate.  As for $y_2$ (or equivalently $\nu=1/y_2$),
Refs.~\onlinecite{PHP-06},\onlinecite{MC-02}, \onlinecite{CF-97} report
$\nu\approx 3$, $\nu=4.0(5)$, and $\nu=2.4(3)$, which are not far from our
much more precise result (but the result of Ref.~\onlinecite{CF-97} is clearly
inconsistent with the quoted errors).  Finally, we quote
$\eta=0.183(3)$,\cite{MC-02} and $\eta=0.1848(3)$, 0.1818(2) (statistical
errors only) obtained in Ref.~\onlinecite{PHP-06} respectively for the $\pm J$
model and for the model with Gaussian distributed couplings.  They are fully
consistent with our result.

\begin{figure*}[tb]
\centerline{\psfig{width=9truecm,angle=0,file=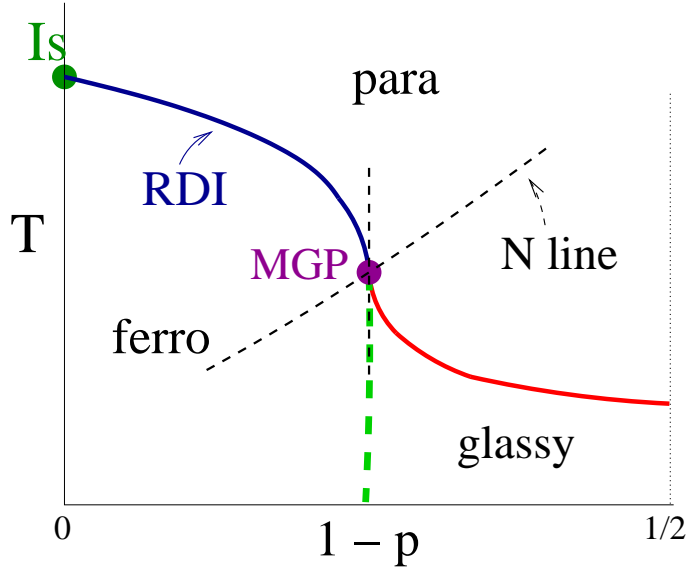}}
\caption{(Color online)
Phase diagram of the 3D $\pm J$ Ising model in the $T$-$p$ plane. 
}
\label{phdiad3}
\end{figure*}

It is interesting to compare the phase diagram of the two-dimensional $\pm J$
Ising model, shown in Fig.~\ref{phdia}, with that of the three-dimensional
$\pm J$ Ising model sketched in Fig.~\ref{phdiad3}.  Recent high-statistics
numerical studies of the $\pm J$ Ising model on a simple cubic lattice have
shown that: (i) the transitions along the PF line
belong to the 3D randomly dilute Ising
(RDI) universality class,\cite{HPPV-07-pmj} with critical
exponents~\cite{HPPV-07} $\nu = 0.683(2)$ and $\eta = 0.036(1)$; (ii) this
line extends up to a magnetic-glassy multicritical point (MGP) located along
the $N$ line, at~\cite{HPPV-07-mcpmj} $p^*=0.76820(4)$, where the relevant RG
dimensions are given by $y_1=1.02(5)$ and $y_2=0.61(2)$ (corresponding to the
thermal and crossover exponents $\nu=1.64(5)$ and $\phi=1.67(10)$); (iii) the
critical behavior along the transition line separating the paramagnetic and
the spin-glass phase is independent of $p$, and belongs to the Ising spin-glass
universality class~\cite{HPV-07} with the correlation-length critical exponent
$\nu=2.53(8)$.

\appendix

\section{Notations}
\label{notations}

The two-point correlation function is defined as
\begin{equation}
G(x) \equiv [ \langle \sigma_0 \,\sigma_x \rangle ],
\label{twof}
\end{equation}
where the angular and the square brackets indicate respectively
the thermal average and the quenched average over disorder.
We define the magnetic susceptibility $\chi\equiv \sum_x G(x)$ and the
correlation length $\xi$
\begin{equation}
\xi^2 \equiv {\widetilde{G}(0) - \widetilde{G}(q_{\rm min}) \over 
          \hat{q}_{\rm min}^2 \widetilde{G}(q_{\rm min}) },
\end{equation}
where $q_{\rm min} \equiv (2\pi/L,0)$, $\hat{q} \equiv 2 \sin q/2$, and
$\widetilde{G}(q)$ is the Fourier transform of $G_1(x)$.  We also consider
quantities that are invariant under RG transformations in the critical limit.
Beside the ratio
\begin{equation}
R_\xi \equiv \xi/L,
\label{rxi}
\end{equation}
we consider the quartic cumulants
\begin{eqnarray}
U_{4}  \equiv { [ \mu_4 ]\over [\mu_2]^{2}}, \qquad
U_{22} \equiv  {[ \mu_2^2 ]-[\mu_2]^2 \over [\mu_2]^2},\qquad
U_d \equiv  U_4 - U_{22},
\nonumber 
\end{eqnarray}
where
\begin{eqnarray}
\mu_{k} \equiv \langle \; ( \sum_x \sigma_x\; )^k \rangle.
\end{eqnarray}
The quantities
$R_\xi$, $U_4$, $U_{22}$, and $U_d$ are
also called phenomenological couplings.
Finally, we consider the derivatives
\begin{equation}
R'_\xi\equiv {d R_\xi\over d\beta},\qquad
U'_4\equiv {d U_4\over d\beta},
\label{derivatives}
\end{equation}
which can be computed by measuring appropriate expectation values
at fixed $\beta$ and $p$.

\end{document}